\documentclass[preprint,proceedings]{rmaa}

\SetYear{2006}
\SetConfTitle{Massive Stars: Fundamental Parameters and Circumstellar Interactions}

\title{The stellar upper mass limit in the solar neighborhood}

\author{
  J. Ma\'{\i}z Apell\'{a}niz\altaffilmark{1}, 
               N. R. Walborn\altaffilmark{2}, 
               N. I. Morrell\altaffilmark{3}, 
                 E. P. Nelan\altaffilmark{2}, 
               V. S. Niemela\altaffilmark{4},
                 P. Benaglia\altaffilmark{4}, 
                 and A. Sota\altaffilmark{2,5}}

\altaffiltext{1}{Instituto de Astrof\'{\i}sica de Andaluc\'{\i}a-CSIC.}
\altaffiltext{2}{Space Telescope Science Institute.}
\altaffiltext{3}{Las Campanas Observatory.}
\altaffiltext{4}{Universidad Nacional de La Plata.}
\altaffiltext{5}{Universidad Aut\'onoma de Madrid.}

\suppressfulladdresses

\listofauthors{J. Ma\'{\i}z Apell\'{a}niz, Nolan R. Walborn, N. I. Morrell, E. P. Nelan, V. S. Niemela, P. Benaglia, and A. Sota}
\indexauthor{Ma\'{\i}z Apell\'{a}niz, J.} 
\indexauthor{Walborn, Nolan R.} 
\indexauthor{Morrell, N. I.} 
\indexauthor{Nelan, E. P.} 
\indexauthor{Niemela, V. S.}
\indexauthor{Benaglia, P.} 
\indexauthor{Sota, A.}

\addkeyword{binaries: close}
\addkeyword{stars: early-type}
\addkeyword{stars: fundamental parameters}
\addkeyword{stars: individual (Pismis 24-1NW, Pismis 24-1SE, Pismis 24-17, HD 93129Aa, HD 93129Ab)}

\begin{document}

\maketitle 

\hyphenation{upper}

\boldabstract{We are using HST GO programs 10205, 10602, and 10898
to test the stellar upper mass limit in the solar vicinity
by attempting to detect optical close companions, thus lowering the calculated
evolutionary masses. We have observed with ACS/HRC all the known 
(as of early 2005) Galactic O2/3/3.5 stars. We also have observations
with HST/FGS and ground-based spectroscopy from LCO and CASLEO. Here we discuss 
our results for Pismis 24 and HD 93129A.}

\section{Pismis 24}

	Pismis 24 is a young cluster ($d=2.5$~kpc, Massey et al. 2001) that 
was thought to contain two very 
early type O stars, Pis 24-1 and Pis 24-17. The zero-age evolutionary mass of Pis 24-1 in
the Walborn et al. (2002) compilation was the largest in their sample, 
210-291~M$_\odot$, well above the currently favored stellar upper mass limit of 
$\sim$150~M$_\odot$.

	Our HRC images resolve Pis 24-1 into two visual components separated 
by 363.86$\pm$0.22~mas. We have used the Baade telescope at LCO to observe
Pismis 24 and MULTISPEC (Ma\'{\i}z Apell\'aniz 2005) to deconvolve the spectra of the
NE and SW components. The three objects (Pis 24-1NE, Pis 24-1SW, and 
Pis 24-17) are of very early type (O3.5 or O4). 

	Our CASLEO spectroscopy shows radial velocity variations of more than
100~km~s$^{-1}$ for the combined spectrum of Pis 24-1NE+SW, 
confirming the earlier finding of Lortet et al. (1984). 
Photometric variability has also been observed (Phil Massey, private communication). 
All of this leads to Pismis 24-1 being composed of three very massive stars, two
of them in an unresolved spectroscopic eclipsing binary (likely, Pis 24-1NE from
the analysis of N\,{\sc iv}~$\lambda$4058).

	The two
Pis 24-1 components show very similar NUV-optical colors, with NE being brighter
by $\sim$0.1~mag. The HRC photometry has been 
combined with 2MASS data and processed with CHORIZOS (Ma\'{\i}z Apell\'aniz 2004) to
accurately measure extinction. 
Pismis 24-1 has $\sim$6~mag of extinction in F550M. Pis 24-17 is more
exinguished by $\sim$0.4~mag.

	With all of the data above and the temperature calibration of Martins 
et al. (2005) we have computed
new zero-age evolutionary masses for Pis 24-1NE (unresolved), Pis 24-1SW, and Pis 24-17. 
The results are 
$\approx$95~M$_\odot$ and uncertainties of 10~M$_\odot$ (the real mass of 
the Pis 24-1NE components should be lower) in all three cases. We conclude
that, although the core of Pismis 24 harbors an unusual concentration of very 
massive stars, none of them threatens to break a stellar upper mass limit of
150~M$_\odot$.

\section{HD 93129A}

	Another target in our sample is Trumpler 14, a young cluster in the
Carina Nebula that contains several very early type O stars, including
HD 93129A, the prototype O2 If* (Walborn et al. 2002). Nelan et al. (2004) 
split HD 93129A into two components using HST/FGS. Later, Ma\'{\i}z Apell\'aniz et al.
(2005) detected a change in the relative position of Aa and Ab with ACS/HRC, 
making this system the only known early-O-type 
astrometric binary.

	We have obtained additional observations of HD 93129A with FGS and HRC and
we have recovered a 1996 FGS observation from the HST archive, yielding data on
the relative position of Aa and Ab over 10 years. The FGS data have been 
reprocessed to take into account the contamination from HD 93129B. The data are 
consistent with a proper motion along the radius vector between Aa and Ab
with an average value of 2.08$\pm$0.23~mas/a. This points towards a highly elliptical and/or 
highly inclined orbit. 

	At this stage we cannot obtain an accurate dynamical mass for the system. 
Assuming a circular orbit with an inclination $\approx$90\arcdeg\
caught at an intermediate apparent separation yields a total mass of 200$\pm$45~M$_\odot$.
We plan to keep observing the system in the following years in order to obtain an accurate
measurement of its total mass and of its mass ratio.

\end{document}